\documentclass[12pt,preprint]{aastex}

\usepackage{unites2e}

\slugcomment{Accepted by the Astrophysical Journal. Tentatively
ApJ {\bf 566} 2002 February 10.}

\shorttitle{Crab above 60 GeV}
\shortauthors{de Naurois, Holder et al.}

\begin{document}


   \title{Measurement of the Crab Flux above 60 GeV with the CELESTE
   \v{C}erenkov Telescope}

   \author{M.~De Naurois\altaffilmark{1,11},
           J.~Holder\altaffilmark{2,12},
           R.~Bazer-Bachi\altaffilmark{3},
           H.~Bergeret\altaffilmark{2},
           P.~Bruel\altaffilmark{1},
           A.~Cordier\altaffilmark{2},
           G.~Debiais\altaffilmark{4},
           J-P.~Dezalay\altaffilmark{3},
           D.~Dumora\altaffilmark{5},
           E.~Durand\altaffilmark{5},
           P.~Eschstruth\altaffilmark{2},
           P.~Espigat\altaffilmark{6},
           B.~Fabre\altaffilmark{4},
           P.~Fleury\altaffilmark{1},
           N.~H\'erault\altaffilmark{4,2},
           M.~Hrabovsky\altaffilmark{8},
           S.~Incerti\altaffilmark{5},
           R.~Le Gallou\altaffilmark{5}, 
           F.~M\"unz\altaffilmark{6,7},
           A.~Musqu\`ere\altaffilmark{3},
           J-F.~Olive\altaffilmark{3},
           E.~Par\'e\altaffilmark{1,10},
           J.~Qu\'ebert\altaffilmark{5}, 
           R.C.~~Rannot\altaffilmark{5,13},
           T.~Reposeur\altaffilmark{5},
           L.~Rob\altaffilmark{7}, 
           P.~Roy\altaffilmark{2}, 
           T.~Sako\altaffilmark{1,14}, 
           P.~Schovanek\altaffilmark{8},
           D.A.~Smith\altaffilmark{5,15},
           P.~Snabre\altaffilmark{9},
           A.~Volte\altaffilmark{6}
         }

\altaffiltext{1}{Laboratoire de Physique Nucl\'eaire des Hautes Energies, 
Ecole Polytechnique, Palaiseau, France}
\altaffiltext{2}{Laboratoire de l'Acc\'elerateur Lin\'eaire, Orsay, France}
\altaffiltext{3}{Centre d'Etudes Spatiales des Rayonnements, Toulouse, France}
\altaffiltext{4}{Groupe de Physique Fondamentale, Universit\'e de Perpignan, 
France}
\altaffiltext{5}{Centre d'Etudes Nucl\'eaires de Bordeaux-Gradignan, France}
\altaffiltext{6}{Physique Corpusculaire et Cosmologie, Coll\`ege de France, 
Paris, France}
\altaffiltext{7}{Nuclear Center, Charles University, Prague, Czech Republic}
\altaffiltext{8}{Joint Lab. Optics, Academy of Sciences and Palacky 
University, Olomouc, Czech Republic}
\altaffiltext{9}{Institut de Science et de G\'enie des Mat\'eriaux et Proc\'ed\'es, 
Odeillo, France}
\altaffiltext{10}{Deceased.}
\altaffiltext{11}{{\em Current address:} Laboratoire de Physique Nucl\'eaire des Hautes Energies,
Universit\'e de Paris VI-VII, France}
\altaffiltext{12}{{\em Current address:} Department of Physics and Astronomy, 
University of Leeds, Leeds, U.K.}
\altaffiltext{13}{{\em Current address:} Nuclear Research Lab,
Bhabha Atomic Research Center, Trombay  Mumbai, India.}
\altaffiltext{14}{{\em Current address:} Solar-Terrestrial Environment 
Laboratory, Nagoya University, Nagoya, Japan}
\altaffiltext{15}{{\em Corresponding Author:} smith@cenbg.in2p3.fr}


\begin{abstract}
We have converted the former solar electrical plant THEMIS 
(French Pyrenees) into an atmospheric \v{C}erenkov detector called CELESTE,
which records gamma rays above $30\U{GeV}$ ($7 \times 10^{24}\U{Hz}$). 
Here we present the first sub-$100\U{GeV}$ detection by a ground based 
telescope of a gamma ray source, the Crab nebula, in the energy region between 
satellite measurements and imaging atmospheric \v{C}erenkov telescopes.
At our analysis threshold energy of $60\pm20\U{GeV}$ we measure a gamma ray rate
of $6.1 \pm 0.8$ per minute. Allowing for 30\% systematic uncertainties 
and a 30\% error on the energy scale yields an integral gamma ray flux of 
$$ I(E>60\U{GeV}) = 
6.2^{+5.3}_{-2.3}\times 10^{-6}\U{photons}\UU{m}{-2} \UU{s}{-1}.$$
The analysis methods used to obtain the gamma ray signal from the raw data are 
detailed. 
In addition, we determine the upper limit for pulsed emission to be <12\% of 
the Crab flux at the 99\% confidence level, in the same energy range.
Our result indicates that if the power law observed by EGRET is attenuated by 
a cutoff of form $e^{-E/E_0}$ then $E_0 < 26\U{GeV}$. This is the lowest energy
probed by a \v{C}erenkov detector and leaves only a narrow range unexplored 
beyond the energy range studied by EGRET.
   \end{abstract}




\section{Introduction}

The Crab was the first source of gamma rays to be convincingly detected  by
ground based telescopes \citep{wee89,vac91} and measurements of its emission spectrum
between $250\U{GeV}$ and $20\U{TeV}$ by various  atmospheric \v{C}erenkov detectors are
now available \citep{whipspec,hegraspec,catspec,masterson}. The flux measurement above
$190 \pm 60\U{GeV}$ recently reported by the STACEE experiment, using the mirrors of a
solar energy research facility to collect \v{C}erenkov light, is the first detection
below 200 GeV by a ground based device \citep{staceecrab}. 
At these energies emission
from the Crab is steady and generally accepted to  come from the nebula, arising from the
inverse Compton scattering of the  synchrotron photons observed at lower energies
\citep{gould,dejager92}.

The EGRET detector on board the Compton Gamma Ray Observatory was used to study the Crab from
$0.03$ to $10\U{GeV}$ \citep{fierro}. The differential energy spectrum measured by EGRET is
well described by the sum of two power laws. Below $0.1\U{GeV}$ the steep spectrum is
attributed to the synchrotron radiation from the nebula, while beyond $0.1\U{GeV}$ the spectrum
hardens and is dominated by pulsed emission. The detailed origin of the pulsar emission is
uncertain. The  outer gap \citep{chengouter,romaniouter,hirotani00} and polar cap 
\citep{hardingpolar} models offer differing pictures. Current very high energy measurements
create difficulties for some outer gap models \citep{whipplepulsed} but refining the picture
requires observations in the heretofore uncovered $10-200\U{GeV}$ region. Determining
the energy at which
pulsed emission is again overtaken by the nebula flux is one of the goals of the
present work.

While the Crab itself is a rather special object, the success of the 
synchrotron self-Compton (SSC) model as applied to the nebula has wide 
implications. On the one hand, this bright source is a test piece for the study
of supernova remnants as the acceleration sites of high energy cosmic rays,
with at issue the question of whether proton or electron acceleration dominates
in a given source. In addition, the SSC mechanism is a cornerstone for the
interpretation of the broadband spectra of AGNs of the blazar class
\citep{Dermer,Ghisellini,Marcowith}.  The experimental data from the Crab which
support the SSC picture consist  of EGRET flux measurements up to $10\U{GeV}$,
with large uncertainties in the  region above $1\U{GeV}$ \citep{egretsteady},
and the extrapolation across more  than a decade in energy to the spectra
measured by the atmospheric  \v{C}erenkov experiments. Clearly, an independent
measurement in the  intervening $50\U{GeV}$ region ($1.2\times 10^{25} \U{Hz}$)
where the  inverse Compton peak in the power spectrum is  expected to lie would
further constrain the parameters of this important model.

The minimum energy threshold, $E_{thresh}$, for current ground based imaging 
atmospheric \v{C}erenkov experiments is limited to $\sim200\U{GeV}$ by the 
rate of accidental triggers due to the night sky light and, in the case of 
single mirror experiments, by the rate of local muon triggers.
The simplest way to reduce the threshold of such an experiment is to increase 
the available mirror area, $A$,  as $E_{thresh}\propto\sqrt{1/A}$; an approach 
which is being followed by the MAGIC collaboration \citep{magic}.
Alternatively, an array of smaller telescopes can be used to reach thresholds 
of $\sim100\U{GeV}$ as predicted for the VERITAS \citep{veritas} and HESS 
\citep{hess} experiments.
These experiments are currently under construction and have not yet started 
taking data. 

CELESTE was designed to reach a very low energy threshold without a large 
expenditure of time and resources by exploiting the mirrors of an existing 
structure; a de-commissioned solar farm in the French Pyrenees. 
An array of 40 such mirrors, used by CELESTE to sample the arrival time and 
photon flux of the \v{C}erenkov wave front at intervals of $\sim30\U{m}$, 
provides a total mirror area of $\sim2000\UU{m}{2}$.
CELESTE uses techniques similar to those pioneered by the early wavefront 
sampling experiments ASGAT \citep{asgat} and THEMISTOCLE \citep{themistocle} 
which operated on the same site, but uses a much greater mirror area and more 
sophisticated trigger logic and data acquisition electronics.
Unlike the imaging experiments, the wavefront sampling method gives no direct 
information about the shower morphology, but alternative methods of hadron 
rejection can be developed using the shape of the wavefront and the 
distribution of \v{C}erenkov light on the ground.
Since their Crab detection cited above, STACEE has lowered
their threshold to $120 \pm 25\U{GeV}$ and expects to descend to $70 \U{GeV}$
\citep{staceeICRC}. The GRAAL experiment also uses a heliostat array but
without secondary optics obtains a relatively high threshold of $250 \pm 110\U{GeV}$
\citep{graal}.

In this paper we present the first measurement of the flux from the Crab above
$60\U{GeV}$, as well as an upper limit for pulsed emission, using the CELESTE 
heliostat array.
We begin with a description of the experiment followed by a summary of 
the data sample and observation techniques.
CELESTE exploits a new experimental technique so we outline the analysis 
method in some detail, including the results of extensive Monte Carlo 
simulations of the detector and the analysis of data taken in common with the 
CAT imaging \v{C}erenkov telescope.
The gamma ray flux measurement and the pulsed flux upper limit are presented 
and the implications for the emission models are discussed.
 Further details on these measurements and on the CELESTE experiment in 
general are available in \citet{thesismdn}. 

\section{The CELESTE Experiment}

\begin{figure}
\epsscale{0.5}
\plotone{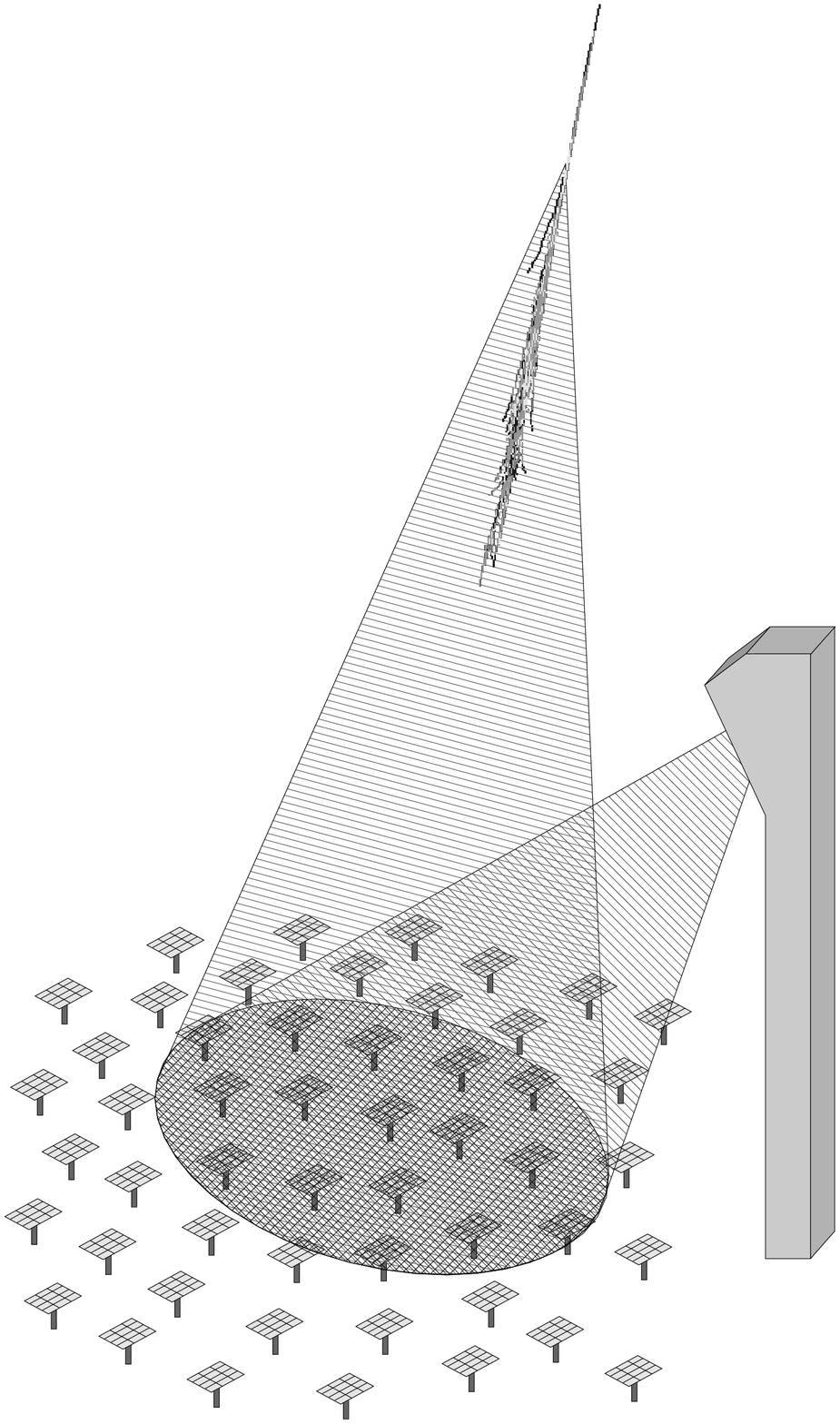}
\caption{Principle of the experimental apparatus.
As the heliostats track a source they reflect Cherenkov light generated by atmospheric
particle cascades to the secondary optics and photomultipliers located near the
top of the 100 meter tall tower. \label{principle}}
\end{figure}

The CELESTE experiment is described in full detail in the experiment proposal 
\citep{proposal} and in \citep{repoman}.
Here we outline the most important features and the status of the experiment 
during the relevant observation period. Fig.~\ref{principle} illustrates the experimental principle.
 
CELESTE uses 40 heliostats of a former solar electrical plant at the Th\'emis 
site in the eastern French Pyr\'en\'ees (N. $42.50^{\circ}$, E. 
$1.97^{\circ}$, altitude $1650\U{m}$).
Each back-silvered heliostat mirror has an area of $54\UU{m}{2}$ and moves on 
an alt-azimuth mount.
The heliostats are controlled from the top of a $100\U{m}$ tall tower, 
located south of the heliostat field, which houses the secondary optics, 
photomultiplier tubes (PMTs) and data acquisition system. 
The alignment of the heliostats has been verified by mapping the images of 
bright stars using the PMT anode current.  

\begin{figure}
\plotone{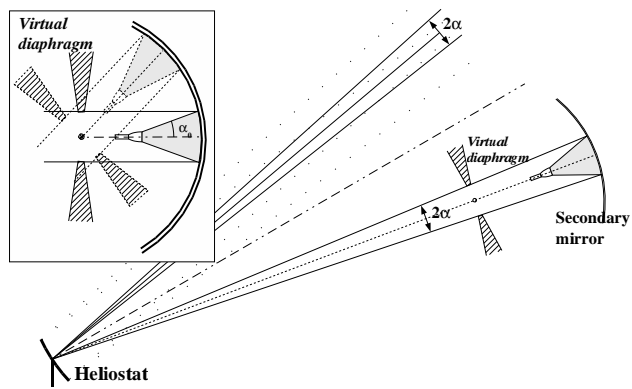}
\caption{The CELESTE secondary optics. Winston cones 
define a ``virtual diaphragm'' which limits the geometrical field of view
to $2\alpha = 10$ mrad (figure not to scale).\label{optics}}
\end{figure}

The light from all 40 heliostats is reflected to the top of the tower. To separate
these signals from each other we use a secondary optical system, as illustrated in
Fig.~\ref{optics}. We have chosen to place the photomultiplier assembly on the optical
axis to minimize coma aberrations, although this results in a loss of light due
to the shadow formed. The spherical mirrors of the secondary optics are divided into
six segments on three levels with three different focal lengths in order to reduce
this shadowing effect and to produce images of approximately the same size regardless
of the heliostat position in the field. One large segment views the farthest
heliostats, two others view those at intermediate distance, and three small segments
are used for the heliostats at the foot of the tower. At the secondary mirror focus is
the entrance face of a solid Winston cone glued to a two-inch PMT (Philips XP2282B),
one for each heliostat. The Winston cone determines the surface area of the
secondary mirror seen by that PMT, such that the optical field-of-view of each tube is
$2\alpha = 10$ mrad (full width). 
This field-of-view is slightly smaller than the angular size of air showers in
our energy range and helps maximze the ratio of \v{C}erenkov to night sky
light.

The single photoelectron (PE) pulse width, after pre-amplifiers 
(gain=100, AC-coupled) and 23 m cables to 
the counting house, is just under $5\U{ns}$ (full width at half maximum). 
PMT gains are set reasonably low ($\sim5\times 10^4$) to avoid damage to the 
tubes from night sky light, and the electronic gains are such that the 
amplitude of a single photoelectron in the counting house is $10\U{mV}$ on 
average. These amplitudes were measured {\em in situ}.
In fact, studies of the average response of each detector to the hadronic 
background events have enabled us to calibrate the relative efficiency of each 
heliostat, and the PMT high voltages are now set so as to correct for this (in 
the range $\sim\pm25\%$) in order to give an even trigger response across the 
heliostat field. 
The PMT signals are sent to both the trigger electronics and to the data 
acquisition system.

\begin{figure}
\plotone{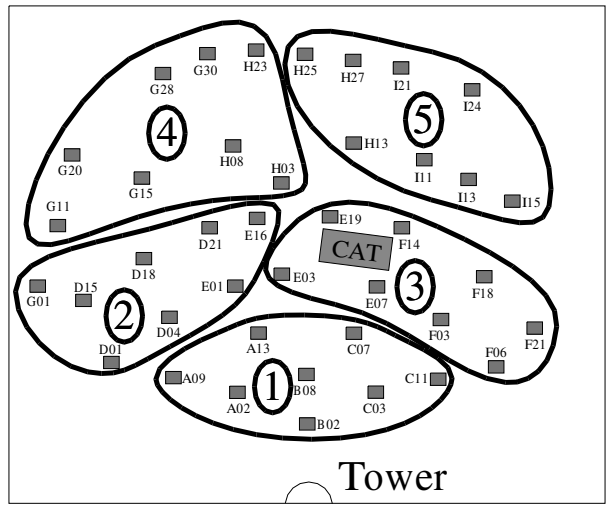}
\caption{The division of the heliostat field into 5 
trigger groups, each of 8 heliostats.\label{trigger}}
\end{figure}

The trigger is designed to reach the lowest possible threshold. Programmable 
analog delays compensate for the changing optical path lengths as the source 
direction changes during the observation. 
The switched-cable delays
broaden the PE pulse widths by a full nanosecond for the maximum delay.
Eight PMT signals are summed in each of five groups as shown in 
Fig.~\ref{trigger}, and the sums enter a discriminator. 
Programmable logic delays further compensate for the varying path lengths between the 
trigger groups.
The logic delay introduces a deadtime of the order of 5\%.
A trigger requires the logic coincidence of at least three of the five groups, 
with an overlap of $10\U{ns}$.
The analog sum over eight heliostats provides us with a good signal to noise 
ratio for the \v{C}erenkov pulse, while the logic coincidence removes triggers 
due to afterpulsing in the PMTs, local muons or low energy hadronic events
illuminating only a few heliostats.

Each PMT signal is further amplified ($\times2$) and sent to an 8-bit Flash 
ADC (FADC) circuit (Etep 301c) that digitizes the signal at a rate of 
$0.94\U{GHz}$ (1.06~ns~per~sample).
The depth of the FADC memory is 2.2~$\mu$s, and one photoelectron corresponds 
to 3 digital counts.
When a trigger occurs, digitization stops and a window of 100 samples centered
at the nominal \v{C}erenkov pulse arrival time is read out via two VME busses 
in parallel. 
Readout requires $7\U{ms}$, which for typical raw trigger rate
of 25 Hz gives an acquisition deadtime fraction of $20\U{\%}$.
The trigger also latches a GPS clock, which is read out and included in the 
data stream.
In parallel with the \v{C}erenkov pulse data acquisition, scalers record the 
single group trigger rates, the final trigger rate, and the readout rate.
Acquisition deadtime is determined from the latter two.
The anode current of each PMT ($\sim10~\mu\mathrm{A}$) is also recorded, as is 
some meteorological information.

\section{Crab Observations }

The observations presented here were taken on clear, moonless nights during 
the Crab season between November 1999 and March 2000.
All the data were taken when the source was within 2.5 hours of transit, that 
is, with an angle from the zenith, $\Theta < 40^{\circ}$.
The observations were made in the ON-OFF tracking mode, in which an 
observation of the source is followed or preceded by an observation at the 
same declination offset in right ascension by an appropriate amount (usually 
20 minutes).
The offset region is then used as a reference to provide a measure of the 
background of cosmic ray events.
It is particularly important in the case of CELESTE to cover the same 
elevation and azimuth ranges during the ON and the OFF source observations as 
the heliostat optical collection efficiencies change appreciably due to the 
projection of the heliostat surface viewed by the PMTs, and (less importantly) 
due to optical aberrations.
Both of these effects depend upon the heliostat orientation and thus upon the 
source direction.
In addition, matching ON and OFF source observations ensures that the ON and 
OFF data were taken using exactly the same path through the delay electronics.

CELESTE has a number of options when deciding how to observe a source.
The majority of the data here were taken in ``single pointing'', wherein all 
the heliostats were aimed at a point $11\U{km}/\cos\Theta$ upward from the 
center of the heliostat field towards the source such that the center of their 
fields of view converged at the expected maximum point of \v{C}erenkov 
emission for gamma showers.
This method collects the largest number of photons, allowing us to operate 
with the lowest possible energy threshold. 
It seems likely, however, that other pointing strategies may provide better 
sampling across the shower and hence better hadron rejection.
With this in mind,  a smaller number of runs were taken using ``double 
pointing'' in which half the heliostats pointed at  $11\U{km}/cos\Theta$, and 
the other half at $25\U{km}/\cos\Theta$. A Monte Carlo study of some different 
pointing methods is available in \citet{thesisnh}. The observing log is 
summarised in Table~\ref{tab:obslog}.

\begin{figure}
\plotone{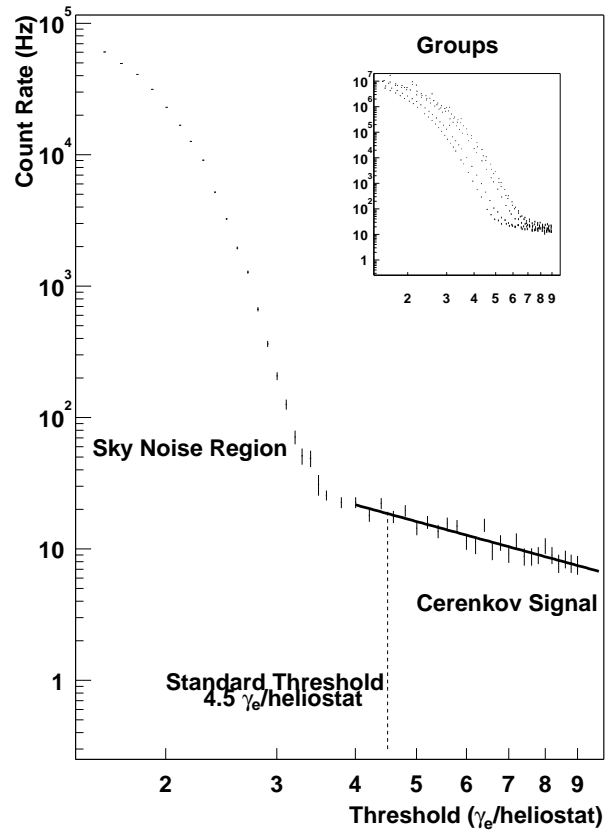}
\caption{The total trigger rate as a function of 
discriminator threshold per heliostat for a particularly clear dark night. 
\textit{Inset:} the rates for each of the 5 trigger groups as a function of 
their discriminator levels, in photoelectrons ($\gamma_e$) per
heliostat.\label{trigrate}}
\end{figure}

The trigger logic was set such that 3 groups out of the 5 were required to 
exceed their discriminator threshold in order to trigger the experiment.
The discriminator threshold levels for each of the 5 trigger groups are checked nightly 
by measuring the trigger rate as a function of 
discriminator level in order to find the break point between accidental coincidences
of random noise pulses and \v{C}erenkov flashes. We set the discriminators such
that the noise triggers contribute less than 1\% of the total rate (Fig.~\ref{trigrate}).
For more than 90\% of the Crab data the discriminator level was set to
$360\U{mV}$, that is, an average of $4.5\U{PE}$ for each of the 8 heliostats
in a group, giving a final trigger rate of $\sim25\U{Hz}$.
Expressing the discriminator level for the analog sum of 8 heliostats in a 
group in terms of PE per heliostat implies that the Cerenkov pulses for each 
heliostat are perfectly in time with each other.
We have checked this timing by reconstructing the group sums using the FADC 
data (although the path to the trigger electronics is not identical to the 
acquisition path) and by oscilloscope measurements during observations.
For the data in this paper, three channels were as much as 2 ns out of time,
while the other 37 channels were less than 1 ns from the average. The three
outlying channels have since been corrected, and
the group sum pulses are now routinely digitized using additional FADCs.

The PMT anode current information and the measured trigger rates of each group 
are very sensitive to changes in the sky conditions and are used to verify 
that the atmosphere was stable throughout the ON-OFF pair.
Any data which showed evidence of poor weather or equipment problems were rejected.
The remaining total data set consists of 14.3 hours of ON-source exposure.

\section{Analysis}

Here we outline the important stages in the analysis of CELESTE data; data 
cleaning, shower reconstruction and hadronic background rejection.
We also present the results of extensive Monte Carlo simulations which have 
been used to derive the analysis techniques and to estimate the sensitivity 
and threshold of the experiment. 
CELESTE has the advantage of being situated on the same site as a well calibrated 
atmospheric \v{C}erenkov imaging telescope, CAT \citep{barrau}.
 This has allowed us to examine the collection efficiency for a subset of the CELESTE 
data and should in the future allow us to cross-calibrate energy, direction,
and acceptance between CAT and CELESTE.

\subsection{Pre-analysis}

For each event which triggers CELESTE, we record a window of $106\U{ns}$ around the
\v{C}erenkov pulse for each PMT using FADCs with a sampling period of $1.06\U{ns}$
(i.e. 100 samples). The FADC window is chosen such that the \v{C}erenkov pulse is
expected to arrive in its center. The beginning of the FADC window (the first 30 
samples) is used to calculate the pedestal level. A small constant voltage offset
applied to the unipolar input of each FADC allows fluctuations in the night sky
background to be measured. Significant differences in the amplitude $\sigma_p$ of
these fluctuations can be seen depending upon the brightness of the region of sky
viewed by the PMT.

The possibility of systematic effects in the data due to differences in night sky
background levels between the ON and OFF source regions of the sky is a  known problem
for atmospheric \v{C}erenkov experiments. \citet{Cawley} proposed a method of
``software padding'', for use with the  Whipple telescope, in which the noise
fluctuations of the ADC signals from the  darker region of sky are artificially
increased to the same level as the  brighter region by adding from a randomly sampled
Gaussian distribution. The effects of night sky light differences can be seen at both
the trigger  level and in parameter distributions during the analysis procedure. These
systematic effects produce a significant difference between the number  of events
remaining from the ON source and OFF source regions after analysis  cuts. The
difference can be either positive or negative, depending on which region  is the
brighter, and a positive difference mimics a real signal. CELESTE is particularly prone
to these problems due to its large mirror area  and angular acceptance per PMT which
combine to give a night sky light  background rate of $\sim1\U{PE}/\U{ns}$. The use of
FADCs introduces another complication in the case of CELESTE: if we  wish to extract
more information than just the integrated charge over the  pulse, a simple addition of
charge sampled from a night sky background  distribution to the measured charge is not
sufficient. The effect of additional sky noise on the complete \v{C}erenkov pulse
shape  must be accounted for. The only way to equalize the night sky background
fluctuations in software  then, is to simulate the response of the PMT-FADC electronics
chain to an  increased rate of single photo-electrons.

We model the single PE pulse using events triggered  by cosmic ray muons passing
through the Winston cones, in standard operating conditions except with the tower door
closed, blocking outside light. These pulses contain many ($\sim50$) PEs, generated at
the photocathode  at almost exactly the same time so to a good approximation the muon
pulse  shape is the same as that of a single PE, only of greater amplitude. The results
agree with those obtained on a test bench with an oscilloscope, and with single PE
pulses measured by the FADC's, using much higher PMT gains which change the PMT time
response somewhat. By simulating the FADC response to single PEs arriving at different
rates, we  obtain a calibration curve of measured fluctuation against the background
rate  of PEs due to night sky light of the form $\sigma_p=s\sqrt{b}$ where $s$ is a 
constant and $b$ is the night sky background rate in $\U{PE}$ per ns.  This curve can
be used to calculate the rate of simulated PEs which needs to  be added to the darker
field in order to equalize the night sky background  fluctuations.

Software padding has been applied to all the ON-OFF pairs used in this analysis but
this alone is not sufficient to remove all the biases caused by night sky background
differences as a brighter region of sky also causes a slight increase in the amount of
near threshold events which trigger the experiment. This can be explained as follows:
additional night sky background fluctuations cause showers which would otherwise be
below threshold to trigger. They also prevent some events which would
otherwise be above threshold from triggering, but because the cosmic ray spectrum is
very steep, the former effect is bigger than the latter, and there is a net
night sky background dependent increase in the number of triggered events. We have
therefore found it necessary to apply a ``software trigger'' at a level higher than
the hardware trigger level, in order to remove these additional small events. Using
the FADC data, we reconstruct the analog sum pulses seen by each of the five trigger
groups and then apply the condition: $\ge$4 groups $>5.0\U{PE}$ per heliostat. This
provides us with comparable background data in the ON and OFF fields, and reduces the
fraction of events triggered by accidental noise coincidences to less than $10^{-3}$,
but has the effect of increasing the energy threshold of the experiment
(Fig.~\ref{ethresh}).

\begin{figure}
\plotone{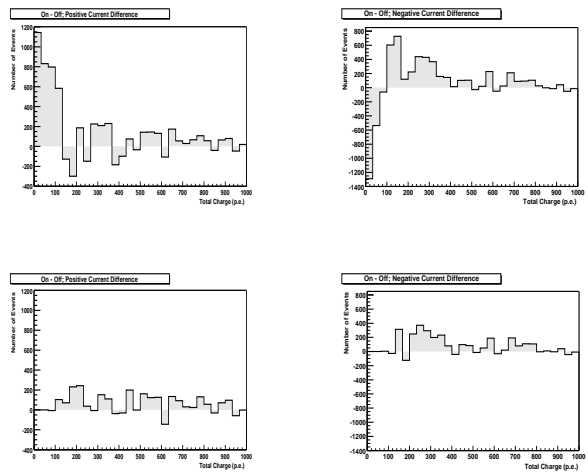}
\caption{Test of the software trigger. The Crab data set 
is divided into two halves as described in the text. The upper plots show the 
ON-OFF distribution of total charge for the raw data after software padding. 
The lower plots show the same after application of the software 
trigger.\label{STtest}}
\end{figure}

To test the performance of the software trigger, we have divided the Crab data 
set into two subsets, based on the sign of the difference in the average PMT 
currents between the ON and OFF source observations.
 Fig.~\ref{STtest} shows the difference between the ON and OFF source 
observations for the distribution of the total charge measured in all the 
\v{C}erenkov pulses for these two subsets.
A clear bias in the number of small events is apparent in the raw data, with 
the direction of the bias depending upon the sign of the current difference. 
After application of the software trigger, the bias has been removed.

\begin{figure}
\plotone{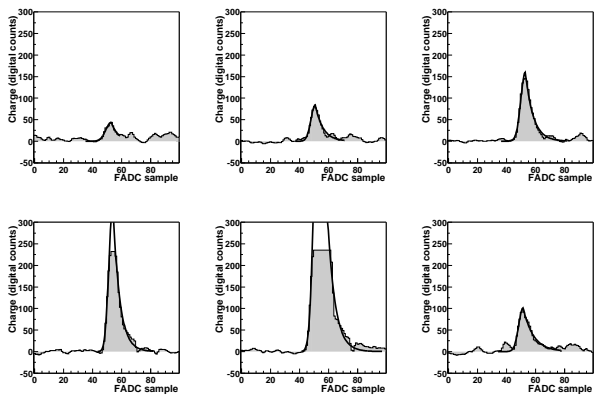}
\caption{An example of the peak finding and fitting 
algorithm for a series of \v{C}erenkov signals of various amplitudes. The 100 
sample FADC data window is plotted in digital counts against FADC sample 
number.\label{peaks}}
\end{figure}

In order to use the information recorded by the FADCs, it is necessary to 
select and parameterize the \v{C}erenkov peaks. This is done by fitting a 
function of the following form:
\begin{equation}
f(t)=\left\{
  \vcenter{\hsize=6cm
  \hbox{\hbox to 3cm{$\displaystyle P e^ 
{\frac{-(t-t_0)^2}{2\sigma_l^2}}$}\quad for \quad $t\le t_0$}
  \vspace{0.5em}
  \hbox{\hbox to 3cm{$\displaystyle P e^ {\frac{-|t-t_0|}{\sigma_r}}$}\quad 
for \quad $t> t_0$}
  \vspace{0.3em}}
\right.
\end{equation}

\noindent where $t$=time in nanoseconds, $t_0$=peak time and $P$=peak amplitude.
Fig.~\ref{peaks} shows some examples of fitted peaks and illustrates the effect of
saturation in the FADCs. Studies using simulated peaks indicate that the peak fitting
algorithm can accurately reconstruct the timing and charge information for peaks which
have saturated the FADCs up to twice their dynamic range. The fit
parameters for each peak are stored for use later in the analysis. Only events having
at least 10 \v{C}erenkov peaks with an amplitude greater than 25 digital counts
($\simeq 8$ P.E.) are used in the analysis ($N_{\mathrm{peaks}}\geq 10$). 

\subsection{Analysis Strategy}

Imaging \v{C}erenkov telescopes have become the most powerful instruments at 
energies greater than $200\U{GeV}$ due to their efficiency in reducing the 
hadronic background. 
Typically, it is possible to reject over 99\% of the background events while 
retaining 50\% of the gamma ray signal \citep{punch}.
At CELESTE energies, a smaller total number of photons and intrinsic 
fluctuations in the shower development mean that the differences between the 
gamma and hadron showers which trigger are less pronounced.
In addition the small field of view of CELESTE, which is necessary to keep the 
night sky light background at a reasonable level, often truncates the shower, 
again causing hadron and gamma showers to look alike.
These points, and also the fact that the trigger system rejects many hadron 
showers at the hardware level, mean that hadron rejection at the analysis 
stage is not very efficient for CELESTE; however, small differences do remain, 
as the gamma showers tend to develop in a more regular manner than the hadron 
showers.

We have written a complete detector simulation package, including a full 
treatment of the complicated optical system of CELESTE and a detailed model of 
the trigger and acquisition electronics, for use with standard air shower 
simulation packages.
Using the Monte Carlo simulations we have investigated various ways 
of exploiting the FADC timing and charge information to provide hadron 
rejection. 
Two rather simple parameters have been studied in detail: the group 
homogeneity, $\sigma_{grp}$, and the shower axis angle, $\theta$.

\begin{figure}
\plotone{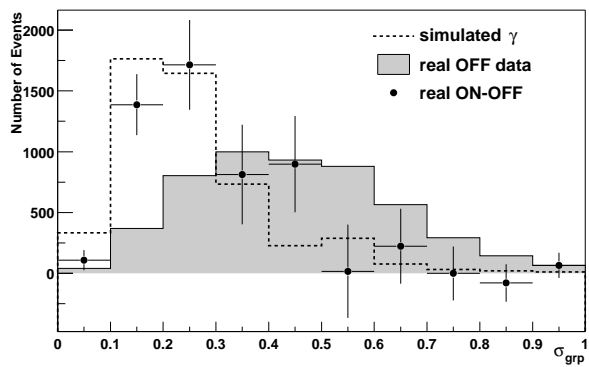}
\caption{The distribution of $\sigma_{grp}$, a measure of 
the homogeneity of the light distribution at the ground, for simulated gamma 
rays, OFF source data and for the difference between the ON and OFF source 
data after the software trigger and with $N_{\mathrm{peaks}}\geq 10$. The 
distributions are normalized to this measured excess.\label{sigmagrp}}
\end{figure}

The group homogeneity is a measure of the homogeneity of the \v{C}erenkov 
light pool at ground level.
It is determined from the variance in the amplitude of the five trigger group 
pulses normalized to the mean amplitude. 
The trigger group pulses are derived by summing the 8 FADC windows of the 
heliostats in each group.

\begin{equation}
\sigma_{\mathrm{\scriptstyle grp}} = \frac{\sqrt{ \big< A_{\mathrm{grp}}^2 
\big> - \big<A_{\mathrm{grp}}\big> ^2 }}{\big<A_{\mathrm{grp}}\big>}
\end{equation}
where $A_{grp}$ are the amplitudes of the 5 reconstructed trigger group pulses.
Fig.~\ref{sigmagrp} shows the distribution of $\sigma_{grp}$ for gamma rays 
and OFF source data after applying the software trigger and requiring a 
minimum of 10 \v{C}erenkov peaks.
The gamma rays were simulated over a range of azimuth and zenith angles so as 
to match the range covered by all the data for the $12.1\U{hour}$ Crab data 
set described in Table~\ref{tab:obslog}. 
The OFF source data shown is the sum of all the OFF source data in this data 
set.
According to this plot, a cut at $\sigma_{\mathrm{\scriptstyle grp}}\leq 0.25$ 
conserves $61\%$ of the gammas which remain after the software trigger, while 
rejecting $85\%$ of the remaining hadrons, giving a quality factor Q=1.6 where:

\begin{equation}
Q = \frac{\hbox{eff}_{\gamma}}{\sqrt{\hbox{eff}_{\hbox{\scriptsize hadrons}}}}
\end{equation}
and $\mathrm{eff}_{\gamma}$ and $\mathrm{eff}_{hadrons}$ are the fraction of 
gammas and hadrons conserved by the cut respectively.

\begin{figure}
\plotone{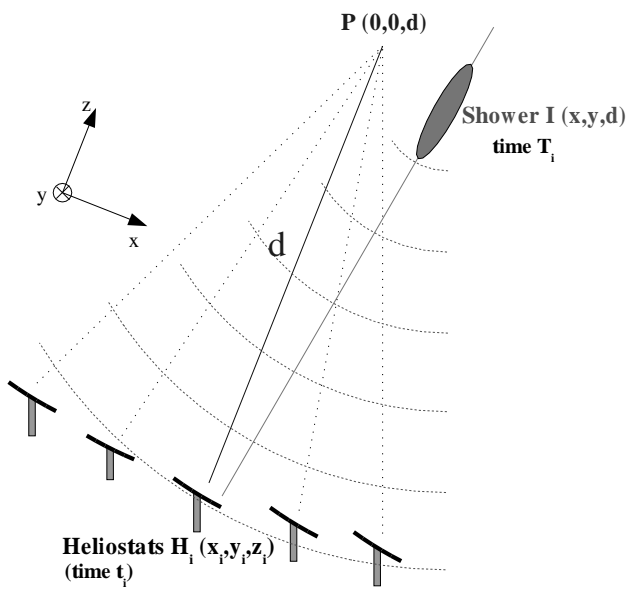}
\caption{An illustration of the shower maximum 
reconstruction. $P$ is the point being tracked, $H_i$ are the heliostat
positions, $I$ is the position to be calculated.\label{spherefit}}
\end{figure}

Low energy gamma ray air showers are only a few kilometres long and the 
majority of the \v{C}erenkov light is emitted from a small region.
The \v{C}erenkov wavefront is therefore spherical to a good approximation 
(Fig.~\ref{spherefit}).
Using the arrival times of the \v{C}erenkov pulses we are able to reconstruct 
this wavefront using an analytical $\chi^{2}$ minimization procedure 
\citep{thesismdn}.

Assuming that the point of emission was at a fixed distance 
$d=11\U{km}/\cos{\Theta}$ from the site towards the source, the fit gives the 
position $I(x,y,d)$ of the shower maximum relative to the tracked point, 
$P(0,0,d)$.
Simulations indicate that this position is reconstructed with an error of 
$\sigma(\sqrt{x^2+y^2})\approx15\U{m}$.

It is important to know the timing resolution for each detector when making the fit.
We have calculated this resolution by studying the response to a nitrogen laser pulse
sent to a diffuser mounted at the top of the tower. The same laser was used for a
similar purpose by the THEMISTOCLE experiment on the same site  \citep{themistocle}.
The timing resolution is also dependent upon the background night sky light level and
on the amplitude of the pulse. This dependency is difficult to test with the laser so
we have measured it by generating simulated peaks, adding them to real night sky
background data and then comparing the reconstructed peak time with the known
injection time of the simulated peak. 
The resolution reaches $\sim0.6$ ns for peaks well
above the night sky noise level, and is worse for larger and smaller peaks due to
FADC saturation and relatively larger night sky fluctuations, respectively.

Using the expected point of maximum emission we can attempt to measure the 
angle, $\theta$, of the shower axis relative to the pointing direction, which 
will be zero in the case of gamma rays originating from a point source at the 
center of the field of view.
To do this we need a second point at ground level, simply calculated by taking 
the mean position of the heliostats on the ground, weighted by the charge 
sampled by each detector. 
More complex algorithms have been tested for calculating the impact parameter, 
but none has proved more effective than this simple method, 
which gives a $1~\sigma$ error of $\sim30\U{m}$ according to the 
simulations.

\begin{figure}
\plotone{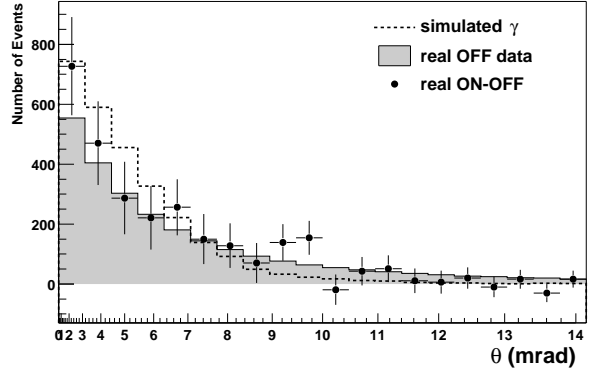}
\caption{The distribution of $\theta$, the shower axis 
angle relative to the source direction, for simulated gamma rays, OFF source 
data and for the difference between the ON and OFF source data after the 
software trigger, with $N_{\mathrm{peaks}}\geq 10$ and 
$\sigma_{\mathrm{grp}}<0.25$.  The distributions are normalized to this 
measured excess. The scale is quadratic.\label{theta}}
\end{figure}

The distribution of $\theta$ for simulated gammas and for real OFF source data 
after the software trigger, requiring a minimum of 10 \v{C}erenkov peaks and 
$\sigma_{grp}<0.25$ is shown in Fig.~\ref{theta}.
As expected, the simulated gamma rays concentrate at small values of $\theta$, 
with an angular resolution of $3.5\U{mrad}$.
 Unfortunately the hadronic background showers, although simulations suggest 
that they can trigger the experiment from as far away as $15\U{mrad}$ from the 
pointing axis, are reconstructed with an angular spread of only 
$\sim4\U{mrad}$.
A cut on $\theta$ alone at $7\U{mrad}$ predicts a quality factor of only 1.1 
after the other cuts have been applied.

In addition to the Crab nebula, CELESTE has recently been used to detect gamma 
ray emission from the TeV blazar Markarian~421 \citep{thesismdn,HeidelbergJH}.
Observations made at the same time by the CAT experiment allowed us to know 
the status of this highly variable source.
The source was observed in December 1999 in a quiescent state and in January 
and February 2000 in an active state, with flares reaching a level of 
$5.5\U{Crab}$ according to CAT. 
The results from the CELESTE analysis show a non-detection for the December 
period (a significance of $-0.3\sigma$ for $1\U{hr}~31\U{min}$ of ON source 
data) and a very significant ($8.1\sigma$ for $5\U{hr}~10\U{min}$) detection 
for the January-Febuary observations. 
These results are noted here as they provide further convincing evidence for 
the stability of the CELESTE analysis.
The ON source star field for the region of Mkn~421 contains a star of 
magnitude 6.1 in the center of the field.
This causes the measured average PMT anode 
currents for the ON source fields to be typically 13\% higher than for the OFF
field. The sky noise differences in the case of the Crab vary by as much as 
$\pm 8\%$ but for all data pairs the dispersion is about $\pm 2\%$,
and the mean difference is smaller than our measurement error.
The non-detection of Mkn~421 in December 1999 implies that the CELESTE 
analysis has correctly dealt with the systematic effects in the data due to 
sky noise differences for this problematic source.
We can therefore be confident that the smaller sky noise differences in the 
case of the Crab observations do not pose a problem, and that our result 
presented in this paper is not significantly biased by systematic 
effects.

\section{\label{sec:results}Results}

Our flux determination uses the results of the analysis of the larger of the two data sets
listed in Table~\ref{tab:obslog}: the $12.1\U{hours}$ of observations with all heliostats
pointing at $11\U{km}/\cos{\Theta}$. The filled circles in figs.~\ref{sigmagrp} and
\ref{theta} show the distribution of the excess events in the ON source data for
$\sigma_{grp}$ and $\theta$ respectively. As predicted for a gamma ray signal, the ON
source excess concentrates at low  values of $\sigma_{grp}$.

Table~\ref{tab:crablow} shows the number of events which remain from the ON and OFF source
observations after the pre-analysis and analysis cuts. As discussed in the previous section,
the first two cuts (the software trigger and  $N_{\mathrm{peaks}}\geq 10$) serve only to
correct for night sky background differences and to ensure that there is enough information to
reconstruct the shower reasonably well. The remaining cuts have been optimized on the
simulations in order to reduce the hadronic background and improve the signal to noise ratio.
As expected from the simulations, the most effective cut parameter is $\sigma_{\mathrm{grp}}$,
with an observed quality factor of 1.4, lower than the predicted 1.6 (quality factors
calculated after the software trigger and $N_{\mathrm{peaks}}$ cuts). We note that at each
stage of the analysis, after the initial pre-analysis  cuts, the ratio of excess to background
increases, from an initial value of  0.6\%, to 5.0\% when all cuts are applied. However, we determine
the Crab flux without using the cut on $\theta$, as the Monte Carlo predicts only a small
improvement in the significance of the result yet adds another source of error into the flux
estimation. 
After the $\sigma_{\mathrm{grp}}$ cut we find an excess of 2727 events,
implying a rate of  $3.8\pm0.5\U{\gamma}\UU{min}{-1}$ and a final statistical significance
of  $7.5\U{\sigma}$.

Table~\ref{tab:EffAll} shows the cut efficiencies at each stage of the analysis procedure
for the OFF source data, for the real Crab ON-OFF source excess, and for the simulated
gamma rays. The agreement between the measured excess and the gamma simulations is 
reasonable, given the large errors on the excess fraction. 

We have also analysed the other set of Crab observations taken in a ``double  pointing''
mode, with half the heliostats pointing at $11\U{km}/\cos{\Theta}$ and the other half at
$25\U{km}/cos{\Theta}$. The results are shown in Table~\ref{tab:crabdouble}. A
statistically significant signal is apparent in this smaller data set, the gamma ray rate
being $4.0 \pm0.8 \U{\gamma}\UU{min}{-1}$ after all cuts, and $5.0 \pm1.0
\U{\gamma}\UU{min}{-1}$ without the $\theta$ cut. The Monte Carlo predicted
an improvement in sensitivity with
this pointing strategy due to its less-biased sampling of 
the \v{C}erenkov light distribution at ground level,
particularly for those showers with large impact parameters. 
In consequence, the cut on
the homogeneity of the light distribution, $\sigma_{\mathrm{grp}}$, becomes more
effective at rejecting the hadronic background. More data is needed to
confirm the double pointing Crab sensitivity of 
$3.4\sigma/\sqrt{hour}$, compared to $2.0\sigma/\sqrt{hour}$ for single 
pointing. Double pointing is now the preferred method of operation for CELESTE. 
Further work is under way in order to determine the optimum pointing 
altitudes, trigger configurations and analysis methods.

\subsection{Detector Sensitivity}

Atmospheric \v{C}erenkov telescopes, unlike satellite experiments, cannot be 
calibrated with a test beam.
Monte Carlo simulations of the detector response to air showers are therefore 
the most important tool for calculating both the detector sensitivity and 
determining the best analysis strategies.
The work presented here has made use of the {\small KASKADE} shower simulation 
package \citep{kaskade}. 
Tests using version 4.5 of the CORSIKA package \citep{corsika} indicate an effective surface 
area for gamma rays $\sim$25\% higher than that of the {\small KASKADE} 
simulations, regardless of the initial photon energy.
The reason for the discrepancy is not yet clear, and an additional systematic 
error has been included in the flux estimation to reflect this. 

\begin{figure}
\plotone{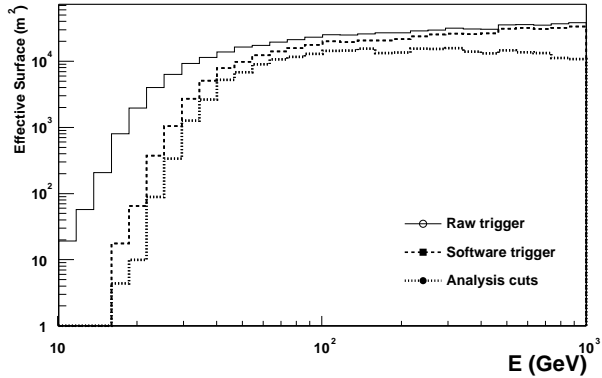}
\caption{The effective surface area for gamma rays of 
CELESTE for a trigger threshold of $4.5\U{PE}$ per heliostat, in the direction
of the Crab at transit. The analysis 
cuts are $N_{\mathrm{peaks}}\geq 10$ and 
$\sigma_{\mathrm{grp}}<0.25$.\label{effsurf}}
\end{figure}

Fig.~\ref{effsurf} shows the effective surface area of CELESTE for gamma rays as
a function of the initial photon energy at the raw trigger level, after the software
trigger, and after the analysis cuts ($N_{\mathrm{peaks}}\geq 10$ and
$\sigma_{\mathrm{grp}}<0.25$), using the {\small KASKADE} Monte Carlo.
The detector simulation was for 11 km single pointing towards
the Crab at transit, with a trigger threshold of $4.5\U{PE}$ per heliostat. 
The curve after cuts can be parametrized as
$A(E) = 14324(1-e^{(15-E)/8.7})^{5.19}(1-e^{(15-E)/23.9})^{2.38}$ $m^2$, with
$E$ in GeV.
The area is an order of magnitude smaller than for an 
imaging telescope because convergent viewing restricts the impact parameter at which a gamma shower 
will be seen by enough heliostats to trigger the experiment.

A valuable partial test of our effective area calculations can be made by  using those
showers which trigger both CAT and CELESTE. Approximately 20\% of the CELESTE events,
corresponding to around 30\% of CAT  events are common and can be identified as such,
with a probability better  than 99.9\%, by their arrival time measured with GPS clocks by
the two experiments. During this observing season we have collected 13 hours of common data
on the  Crab. The standard CAT analysis \citep{lebohec} when applied to the full data 
set results in an excess of 1268 gamma events over a background of 3131  hadrons. The
same analysis applied only to the common events produces an excess of 418 gammas over a
background of 526 hadrons. From these numbers we see that imposing a CELESTE trigger
increases the  signal to noise ratio in the CAT data sample by a factor of two, although
it does not improve the significance of the result as the data sample is smaller.

\begin{figure}[t]
\plotone{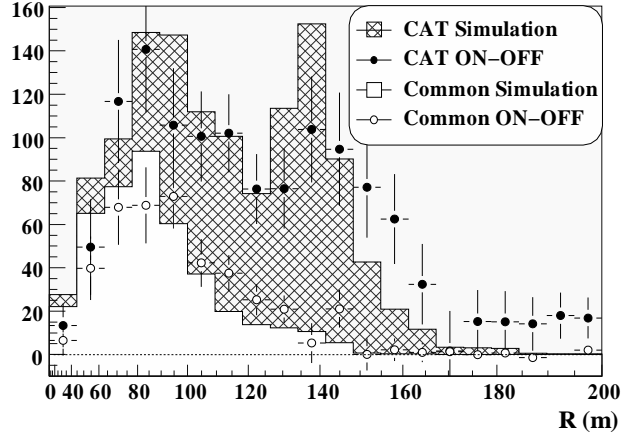}
\caption{The impact parameter (R) distribution as 
measured by the CAT experiment for real and simulated gamma rays. The hatched 
histogram (and filled circles) includes all the CAT events, the clear 
histogram (and open circles) includes only those events seen by both CAT and 
CELESTE.\label{commonsurf}}
\end{figure}

CAT measures the shower impact parameter with   better resolution than CELESTE
\citep{lebohec}. Fig.~\ref{commonsurf} shows  this reconstructed impact parameter for
simulated data, and for the excess  events from the common Crab data set. The data are
well reproduced by the simulations in terms of both the shape of the distributions and in
the predicted fraction of common events.  This gives us confidence that the effective
surface area for CELESTE, at least in the energy region of the common CAT-CELESTE events,
is understood.

\begin{figure}[t]
\plotone{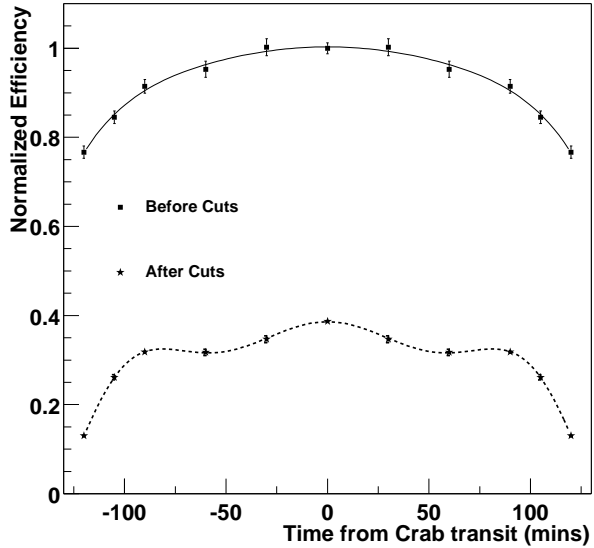}
\caption{The simulated gamma ray detection efficiency for 
CELESTE Crab observations as a function of time of the observation relative to 
Crab transit. The efficiency is normalized to the raw trigger simulated gamma ray 
rate at the Crab transit. The simulations were made only for azimuth directions
before Crab transit - we rely on the symmetry of the heliostat field to extrapolate
to after transit. The errors are statistical only.
The analysis cuts are after the software trigger, 
with $N_{\mathrm{peaks}}\geq 10$ and 
$\sigma_{\mathrm{grp}}<0.25$.\label{efficiency}}
\end{figure}

\begin{figure}
\plotone{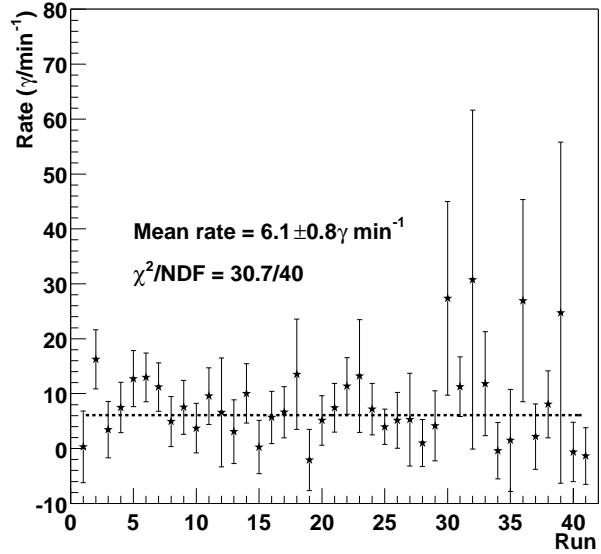}
\caption{The measured flux for each ON-OFF pair for the 
41 Crab runs. The points on the right of the plot have large errors bars as 
these runs were taken towards the end of the season with the source often far 
from transit.\label{SigmaCrab}}
\end{figure}

The effective area varies with the
source position in the sky, as indicated in Fig.~\ref{efficiency}. Knowing the azimuth
angles under which the Crab was observed, we have used the polynomial fit in
Fig.~\ref{efficiency} to correct our measured gamma ray rate. In
addition, for each run we correct for our acquisition dead time of $\sim20\%$
which is measured during the observations. There is no evidence for time variability in
the measured flux of high energy emission from the Crab nebula for gamma ray energies
above and below the energy range of CELESTE \citep{egretsteady,vac91}.
Fig.~\ref{SigmaCrab} shows the rate calculated for each of the 41 ON-OFF  pairs, after
accounting for the varying gamma ray detection efficiency. A constant fit to these points
has a positive mean and a $\chi^{2}$ value of 30.7 for 40 degrees of freedom, as would be
expected for a steady signal, which gives us further confidence in the stability of
the CELESTE analysis. We obtain the
corrected measurement of $6.1\pm0.8\U{\gamma}\UU{min}{-1}$ (statistical
uncertainty only).

Knowing the effective surface area as a function of energy we can calculate the expected
response of CELESTE to a typical spectrum of gamma rays. Fig.~\ref{ethresh} shows the
energy distributions of simulated events for an input $E^{-2}$ differential gamma ray
spectrum, close to the spectral shape for high energy emission from
the Crab in the CELESTE energy range \citep{whipspec}. A useful definition of the energy
threshold for atmospheric \v{C}erenkov detectors is the energy at which the differential
gamma ray rate is maximum for a typical source. According to this definition, the energy
threshold for CELESTE at the raw  trigger level for a source at the position of the Crab
at transit is $\sim30\U{GeV}$ 
\footnote {The Themis solar plant was designed to collect sunlight and
is most efficient when pointing towards the south at an angle of $20^{\circ}$ from the
zenith, which is the same position as for the Crab at transit.}. 
The gamma rays have been simulated with the same distribution of
azimuth and zenith angles as the 11 km Crab observations, increasing the energy threshold 
to $\sim40\U{GeV}$ at the raw trigger stage. As mentioned in the previous section, a
software trigger is applied during the analysis to correct for night sky background
effects in the data. This increases the energy threshold to a level of $\sim60\U{GeV}$.
Further analysis cuts ($N_{\mathrm{peaks}}\geq 10$ and $\sigma_{\mathrm{grp}}<0.25$)
reduce the number of gamma rays observed, but do not increase the energy threshold. 

The systematic errors on our measurement have two different origins. 
The uncertainty on the energy scale is due principally to errors 
in the conversion of the measured signal to a flux of \v{C}erenkov photons, 
which is a combination of many factors 
(photon losses through the optical system, PMT quantum
efficiencies, electronic calibration errors). 
We bracket the overall uncertainty arising from the combination of these
elements as follows.
First, during the CELESTE prototype studies we measured the night sky background
in our wavelength range at Themis to be 
$(2.3 \pm 0.4)\times 10^{12} \U{photons}\U{m}^{-2}\U{sr}^{-1}\U{s}^{-1}$
\citep{berrie}, in the direction of the Crab at transit 
(20 degrees south of zenith, towards the populated valley below the site).
From this we expect 1 photoelectron per nanosecond per phototube,
corresponding to anode currents of $8\mu$A, 
close to the observed range around $10\mu$A. 
Studying the FADC pedestal widths used in the padding software also yields
values of $\sim 1$ photoelectron per nanosecond per phototube.
We further compare currents measured while aligning the heliostats
using star scans with predictions from the optical simulation: 
the measured values are typically 20 \% less than expected. Results of
studies of the atmospheric extinction using CCD photometry and a LIDAR
will be reported in future work but are not included in the present study.
Finally, the observed cosmic ray trigger rate is 30\% higher than predicted
by the Monte Carlo. From these considerations we believe the energy scale
uncertainty to be less than $\pm 30$\%. 
The corresponding acceptance curves are $A( (1 \pm 0.3)E)$, leading to an
uncertainty on our threshold. The input spectrum assumed in determining
the absolute flux (see discussion below) has little effect on the threshold.
For this analysis then, we quote an energy threshold of $60\pm20\U{GeV}$.

The other principal source of systematic error is the uncertainty on our efficiency
for detecting gamma rays. 
As mentioned earlier, there is an energy independent discrepancy of 
25\% in the effective surface area as calculated using two different 
shower generation Monte Carlos. Both Monte Carlo's use the U.S. standard atmosphere.
\citet{bernlohr} recently explored the effects of different atmospheric profiles
on the Cherenkov light yield, finding a $\pm 10\%$ variation for 100 GeV gamma rays
for midlatitude summer and winter atmospheres, neglecting aerosol variations.
We conclude that the uncertainty in the Cherenkov light yield for gamma rays in
our energy range is 25\%.
We also assign a systematic error of 10\% to the cut efficiencies
deduced from the simulations (Table 3).

\subsection{Flux Estimation}

At present the event-by-event energy determination in CELESTE is poor.
To compare our rate measurement with models and with results from other
experiments requires convoluting our detector acceptance, $A(E)$ (see Fig.~\ref{effsurf}),
with an assumed source spectrum. 
The simplest hypothesis is that of a power law differential flux, $1/E^\gamma$.
In a $\nu F_\nu$ representation (or, equivalently, $ E^2 {dN \over dE}$) 
the CELESTE energy range corresponds to the top of the parabola-like 
spectral shape attributed to inverse Compton
production of gamma rays in the nebula \citep{whipspec}, and $\gamma=2$ is
a good approximation. It yields an integral result for CELESTE
of $I(E>60\U{GeV}) = 5.5\times 10^{-6} \U{photons}\UU{m}{-2} \UU{s}{-1}$.
STACEE used this approach, with $\gamma=2.4$ \citep{staceecrab}.

A more realistic hypothesis recognizes that the spectrum deviates
from a pure power law. 
We use a parabola-like spectral shape of the form
$$ E^2 {dN \over dE} = kE^{\alpha+\beta\log_{10}E}.$$
Above $500\U{GeV}$ we use the values of $k,\alpha$, and $\beta$ taken from  
CAT \citep{masterson}. Below $500\U{GeV}$ we let $\beta$ be a free parameter, 
but require continuity at 500 GeV. We determine $\beta_C$
such that the convolution with $A(E)$ yields our measured rate and
thus obtain an integral flux of
$I(E>60\U{GeV}) = 6.2\times 10^{-6} \U{photons}\UU{m}{-2} \UU{s}{-1}$.
We further vary $\beta$ and determine that the range of 
($3.9$ to $11.5)\times 10^{-6}\U{photons}\UU{m}{-2} \UU{s}{-1}$ is consistent
with the rate and acceptance uncertainties, including that of the energy scale.
Repeating the process using the Whipple \citep{whipspec}, 
or HEGRA \citep{hegraspec} spectra gives very nearly the same results.
We thus determine our flux to be 
$$ I(E>60\U{GeV}) = 
6.2^{+5.3}_{-2.3}\times 10^{-6}\U{photons}\UU{m}{-2} \UU{s}{-1}.$$
We applied this procedure above 190 GeV to compare with STACEE and obtain 
$I(E>190\U{GeV}) = 1.8\times 10^{-6} \U{photons}\UU{m}{-2} \UU{s}{-1}$
in agreement with their result of
$I(E>190\U{GeV}) = 2.2\pm 0.6\pm 0.2 \times 10^{-6} \U{photons}\UU{m}{-2} \UU{s}{-1}$.

\begin{figure}
\plotone{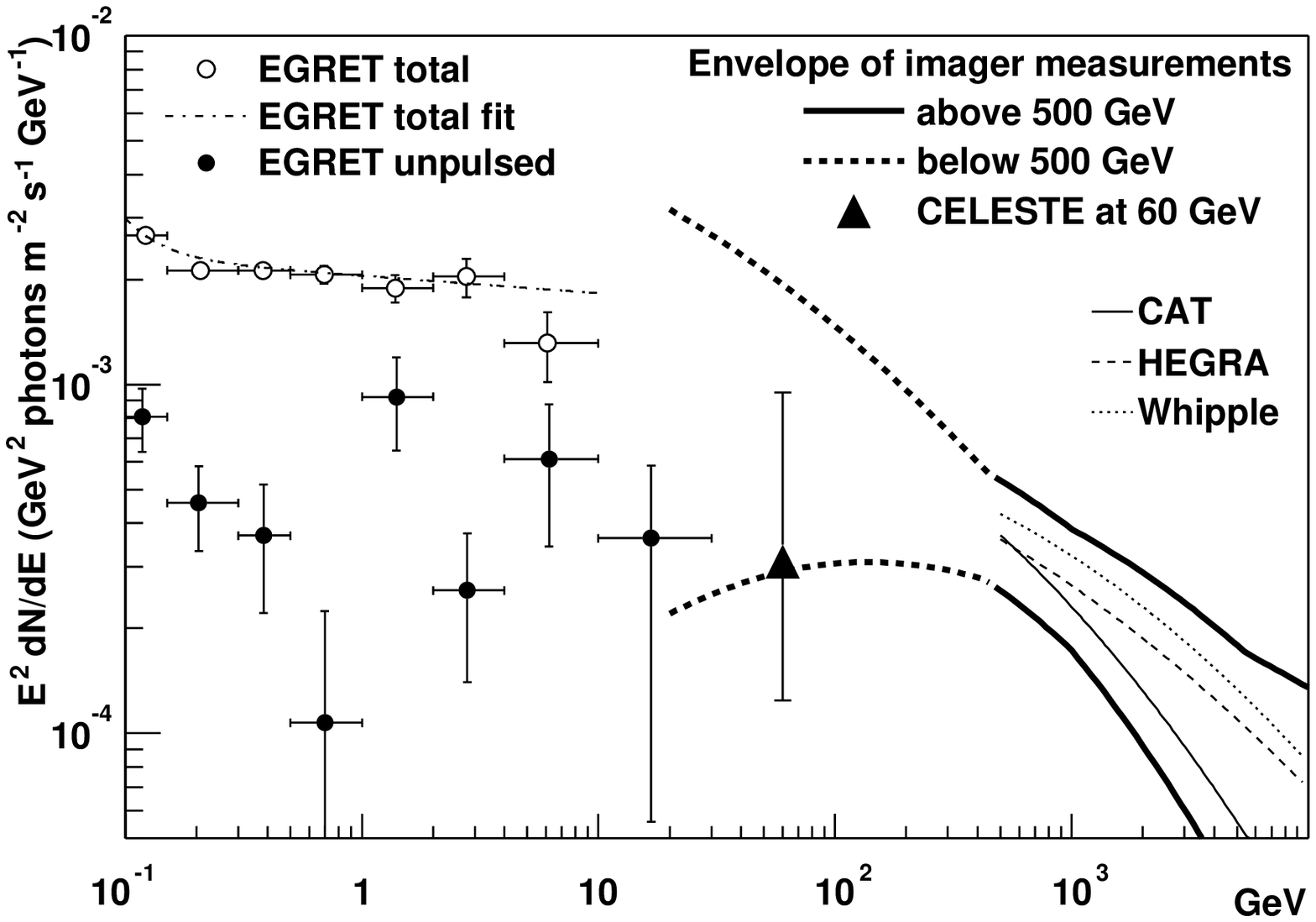}
\caption{The very high energy spectrum of the Crab.
The spectra measured by three Cherenkov imaging telescopes are shown
by the thin curves \citep{whipspec,hegraspec,masterson}. 
Varying the imager fit parameters by one standard deviation gives the range bound by
the solid thick curves.
The thick dashed curves extend the envelope to lower energies.
The flux shown for CELESTE (triangle) corresponds to the spectral
shape which, when convoluted with the detector acceptance, yields
the observed gamma ray rate (see text). 
The point is placed at the nominal
energy threshold, with the energy scale uncertainties included in
the error in the flux determination. 
Also shown is the spectrum of all photons detected by EGRET (open dots) \citep{fierro}, 
and the EGRET data attributed to the nebula (black dots) \citep{egretsteady}.
\label{spectrum}}
\end{figure}

To represent this integral measurement on a differential plot, 
we use $\beta_C$ to calculate $ E^2 {dN \over dE}$ at our energy
threshold. This is shown as a triangle in Fig.~\ref{spectrum}.
As above, the error bar is obtained by finding the range of $\beta$ 
that accomodates the uncertainties on our measurement.
The value shown is 
$ 3.1^{+6.3}_{-1.8}\times 10^{-4} \U{GeV}\U{m}^{-2} \U{s}^{-1}$.
Fig.~\ref{spectrum} also shows the imager measurements, as well as
the envelope defined by varying the imager fit parameters
$k,\alpha$, and $\beta$ by one standard deviation around
their central values. Our measurement favors
the lower part of the range allowed by the imagers and is compatible
with the results from EGRET.




\section{Periodicity Search}
 
One of the primary goals of the CELESTE experiment is to investigate the 
periodic emission from gamma ray pulsars in the cutoff region below 100\U{GeV}.
The CELESTE data include the arrival time of each event measured to a 
precision of $< 1\mu$s using a time-frequency processor slaved to a Global 
Positioning System (GPS) clock which provides synchronisation every second.
This timing information has been used to search for evidence of periodicity in 
our Crab data.

In order to verify our periodic analysis procedure we have made observations 
of the optical emission from the Crab pulsar using the CELESTE heliostats.
Given the optical flux from the Crab pulsar \citep{percival}, we expect a flux 
of $\sim1\times10^4 \U{PE}\UU{s}{-1}\UU{heliostat}{-1}$ over a night sky 
background of typically $\sim1\times10^9 \U{PE}\UU{s}{-1}\UU{heliostat}{-1}$.

\begin{figure}[t]
\plotone{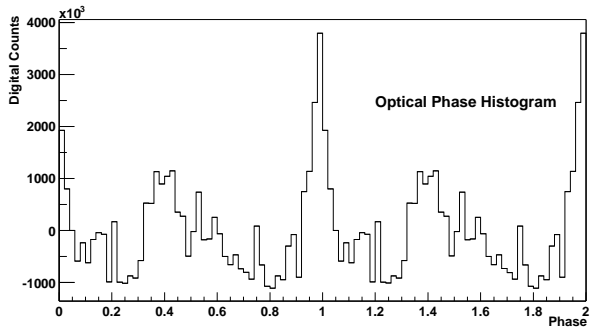}
\caption{Phase histogram for the optical Crab 
data.\label{optical}}
\end{figure}

In standard operation, the PMT anode currents for all forty heliostats are 
converted to a buffered voltage which is digitized and stored with the data 
stream. The current-to-voltage conversion integrates the signal over $<1~\U{ms}$.
For the optical pulsar study, three of these current outputs were AC-coupled, in
order to subtract the steady component due to the night sky background and the
nebula, and sent to a 16-bit ADC card readout by a PC at a frequency of 
$2000\U{Hz}$. A GPS time reference was obtained for the optical data by sending
the same pulse every $10\U{s}$ as a trigger to CELESTE and as data to the ADC
card. We then tracked the Crab pulsar and recorded the current fluctuations
during 30 minutes. The synchronised times were converted to the solar system
barycenter frame using the JPL DE200 ephemeris \citep{bary}. 
Fig.~\ref{optical} shows the phase histogram using the frequency ephemerides 
obtained by \citep{LynPritRob}. The double-peaked signal from the pulsar is
clearly visible. We use the same code to calculate the phase of the air shower
events. 

From the EGRET pulsar detections, the TeV upper limits, and the model 
predictions it is clear that the search for pulsed gamma ray emission requires 
as low an energy threshold as possible.
To date we have no evidence of a pulsed signal.
We present the pulsar search using the same analysis as used to measure the 
steady emission flux, that is, applying the software trigger, 
$N_{\mathrm{peaks}}\geq 10$ and $\sigma_{\mathrm{grp}}<0.25$. 
Although this raises our energy threshold, we take this cautious approach 
because the efficiency is better understood.

\begin{figure}[t]
\plotone{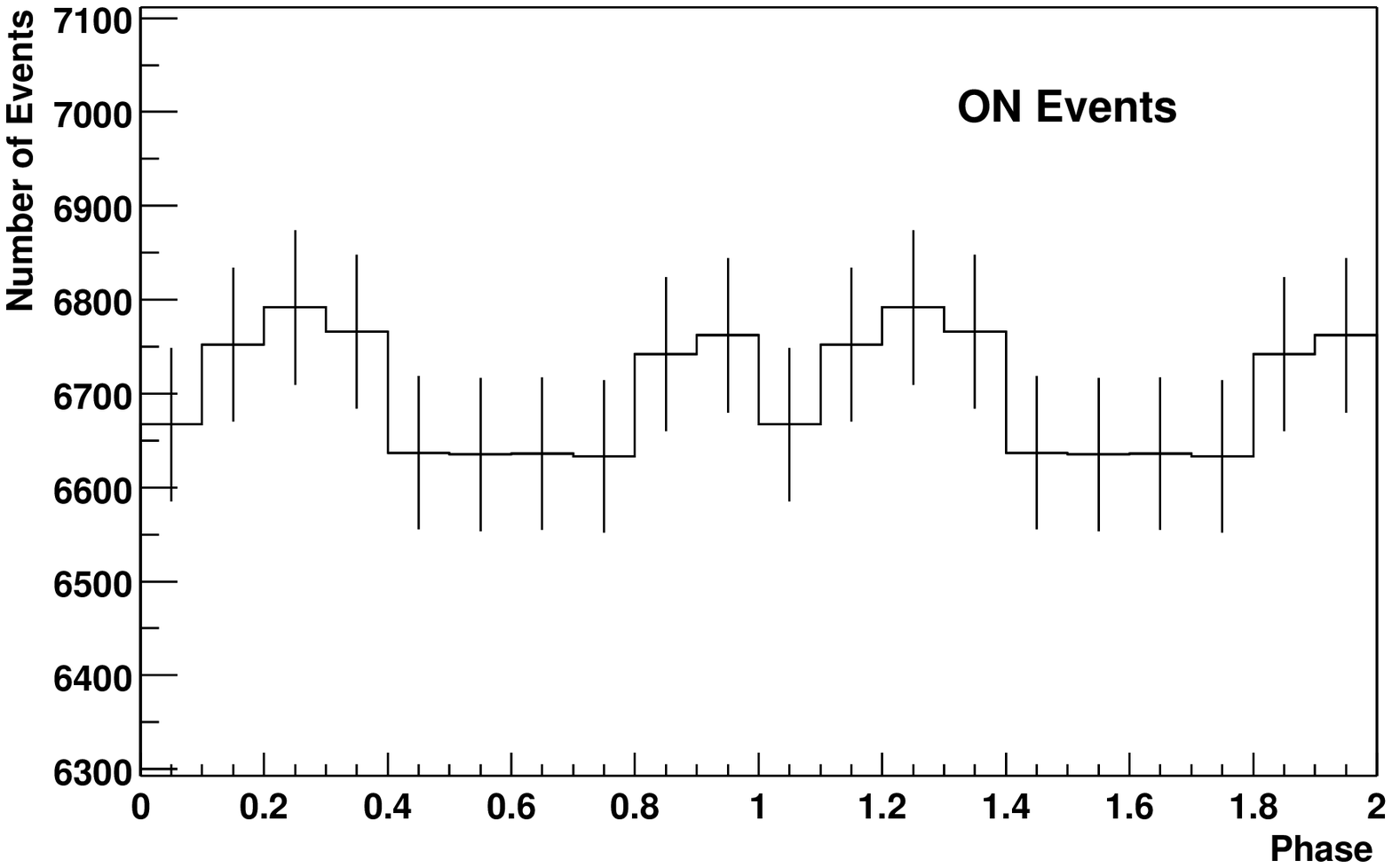}
\plotone{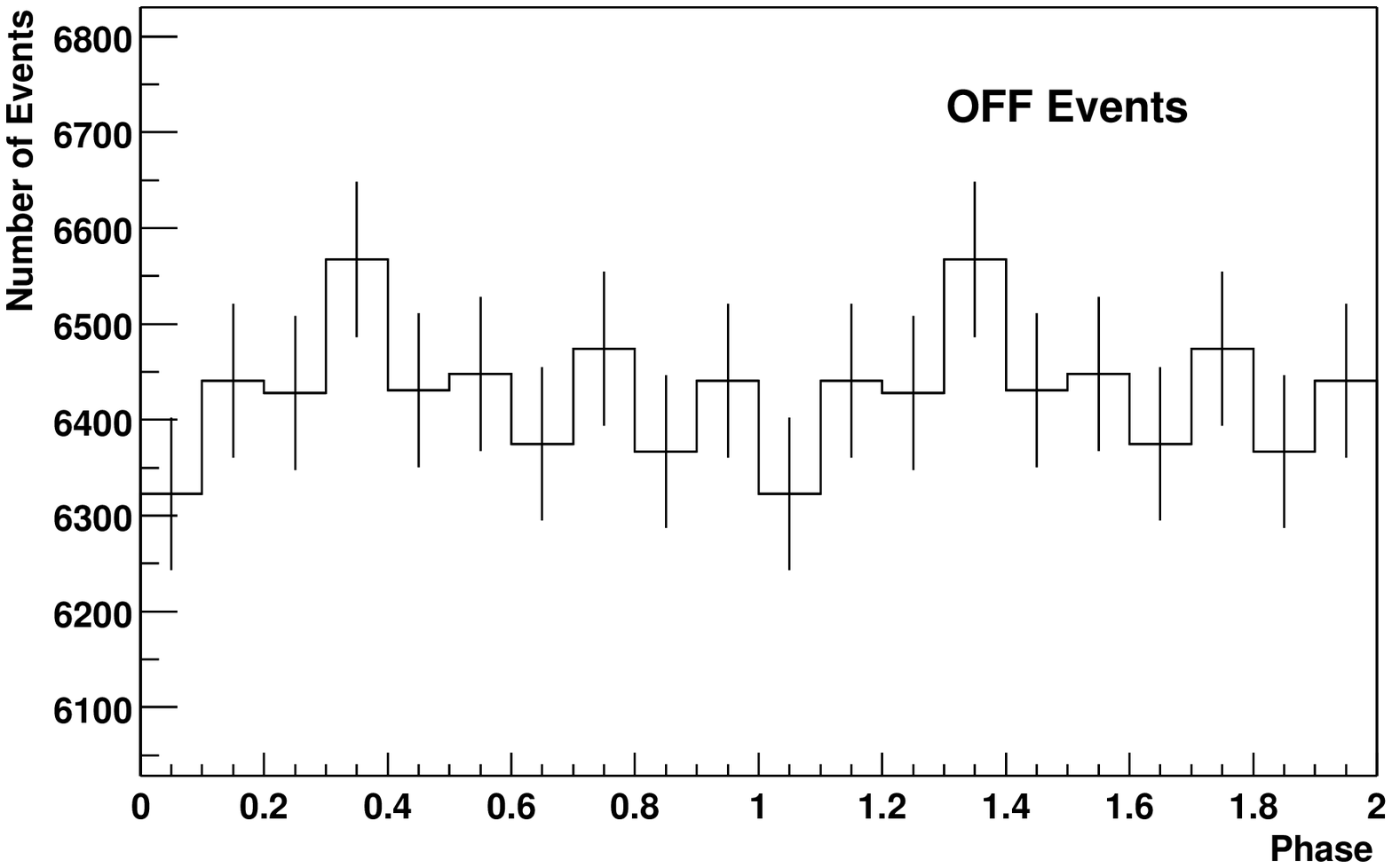}
\caption{Phase Histograms for the ON 
source and OFF source CELESTE 11km Crab observations. Both are statistically 
flat.\label{phasograms}}
\end{figure}

The light curves of both the ON source data and the OFF source data
remaining after cuts for the $12.1\U{hour}$ Crab data set of 
Table~\ref{tab:obslog} are shown in Fig.~\ref{phasograms}. 
Table~\ref{tab:pulsed}
summarises the contents of the plots as well as the results of the H-test 
\citep{deJager94}.
 The distributions are statistically flat.
In order to calculate upper limits for the pulsed emission we assume that the 
pulse profile is the same as that seen by EGRET at lower energies with 
emission concentrated in a main pulse in the phase range 0.94-0.04 
and a secondary pulse in the range 0.32-0.43 \citep{fierro}.
We use the method of Helene to determine an upper limit of $< N_p = 332$ 
pulsed events
at the 99\% confidence level \citep{helene}. This corresponds to 12\%
of the observed steady signal.

\begin{figure}
\plotone{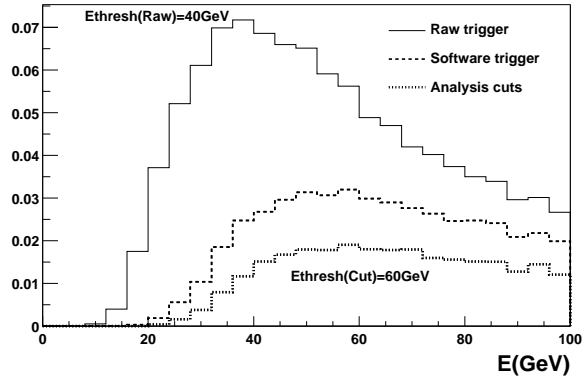}
\caption{The simulated response of CELESTE to an $E^{-2}$ 
power law gamma ray spectrum (normalized to the integral of the raw trigger 
curve). The gamma rays have been simulated with the same distribution of 
azimuth and zenith angles as the 11 km Crab observations. The analysis cuts are 
$N_{\mathrm{peaks}}\geq 10$ and $\sigma_{\mathrm{grp}}<0.25$.\label{ethresh}}
\end{figure}

We include the detector acceptance as follows.
We take the double power law fit of the total spectrum measured by 
EGRET \citep{fierro},
and attenuate the sum with an exponential cutoff,
$$ {dN \over dE}|_{att} = 
[0.7(E/100)^{-4.89} + 2.3(E/100)^{-2.05}]e^{-E/E_0},$$
in units of $10^{-8}$ photons cm$^{-2}$s$^{-1}$MeV$^{-1}$.
We convolute this spectrum with the acceptance after cuts shown in 
Fig.~\ref{ethresh}, and find that for $E_0=20$ GeV we would expect $N_p$ events.
Including the 30\% uncertainty in the energy determination 
degrades this value to  $E_0=26$ GeV. Fig.~\ref{pulsedspec} shows ${dN \over 
dE}|_{att}$, where we have placed a point at the energy threshold obtained for 
our steady signal to guide the eye.
 We note that our limit is not directly comparable to that obtained by the 
STACEE \citep{staceecrab} group since they used the larger acceptance 
corresponding to their measured steady spectrum for comparison with the 
prediction of TeV pulsed emission.
 Our hypothesis of an attenuated EGRET spectrum restricts our acceptance to 
the low energy range of Fig.\ref{ethresh}, yet our upper limit still provides 
the most constraining measurement so far on the position of the cutoff point.
 In the future, improved trigger electronics and observing and analysis 
strategies optimized for pulsar observations should allow us to increase our 
acceptance at low energy.

\section{Discussion}

The radiation from the Crab nebula is dominated by non-thermal emission which 
is believed to be generated by synchrotron radiation from highly relativistic 
electrons with energies up to $\sim10^{15}\U{eV}$.

The electrons are accelerated at the shock front where a relativistic wind of 
charged particles emerging from the pulsar meets the surrounding nebula 
\citep{rees,kennel}.
 Recent high resolution X-ray observations by the Chandra observatory have 
shown an inner ring of X-ray emission which may correspond to the position of 
this shock \citep{chandra}. 
\citet{ahaato} and \citet{atoaha} have described the electrons in terms of two 
populations of different energies.
The first, generated over the whole lifetime of the nebula and covering 
energies up to $\sim100\U{GeV}$, produces synchrotron radiation from radio 
wavelengths to the far infra-red while the second, more recently accelerated 
population, with energies $>10^{12}\U{eV}$ produces synchrotron emission from 
the infra-red up to $\sim$1 GeV.

It was first suggested by \citet{gould} (also \citet{rieke} and 
\citet{grindlay}) that the synchrotron self-Compton mechanism could give rise 
to radiation from the Crab above $1\U{GeV}$.
This process, in which inverse Compton scattering of the synchrotron photons 
by the relativistic electrons boosts the photons up to much higher energies, 
has been modelled by various workers, most recently \citet{dejager92}, 
\citet{atoaha} and \citet{whipspec}.
While the synchrotron photons are the most important component, photons due to 
infra-red emission from dust and to the microwave background will also be 
upscattered and contribute significantly to the high energy emission.

Fig.~\ref{spectrum} shows the result of this work along with the measurements 
from EGRET and three atmospheric Cherenkov imaging telescopes.
The shape of the inverse Compton spectrum is relatively insensitive to the 
model parameters, but the absolute flux depends strongly upon the magnetic 
field strength in the emitting region, which in turn depends upon $\sigma$, 
the ratio of the magnetic field strength to particle energy density in the 
pulsar wind.
\citet{atoaha} have proposed that an additional component due to 
Bremsstrahlung radiation from the relativistic electrons in dense filaments of 
nebular gas may provide an increased flux in the $1-100\U{GeV}$ range, which 
could account for a possible discrepancy between the models and the EGRET 
points around $1\U{GeV}$.
The uncertainties are still large but the CELESTE measurement does not 
seem to point towards such an effect.
The calibration of such a complex instrument as CELESTE is a large project in 
itself. Our measurement errors are currently dominated by systematic effects which 
should decrease as this work proceeds, the most important being to
improve our determination of the energy scale.

The Crab pulsar is a source of 33\U{ms} pulsed radiation from radio  wavelengths to GeV
gamma ray energies. Periodic emission is observed by EGRET up to energies of $10\U{GeV}$ 
\citep{ramana}. Despite early claims \citep{gibson,bhat,dowthwaite}, no pulsed emission
has been detected by the present generation of ground based atmospheric  \v{C}erenkov
experiments. The previous best upper limits are at $250\U{GeV}$, from the Whipple
\citep{whipplepulsed} and CAT \citep{musquere} groups, and the limit at $190\U{GeV}$
by STACEE \citep{staceecrab}.

Two general classes of models have been proposed to describe the 
pulsed gamma ray emission from the high energy pulsars observed by EGRET. 
In the polar cap models \citep{hardingpolar, hardingpolar96, Sturner}  
electrons accelerated from the neutron star surface at the magnetic pole 
emit by curvature radiation or magnetic inverse Compton scattering, 
triggering photon-pair cascades in the pulsar magnetosphere from which 
the observed radiation emerges.
Outer gap \citep{chengouter,romaniouter} models place the emission region 
in the outer magnetosphere where electrons are accelerated across charge 
depleted regions near the light cylinder.
Both models predict a cutoff in the pulsed emission below $100\U{GeV}$, and 
the exact position for the cutoff can be used to discriminate between them.

\begin{figure}
\plotone{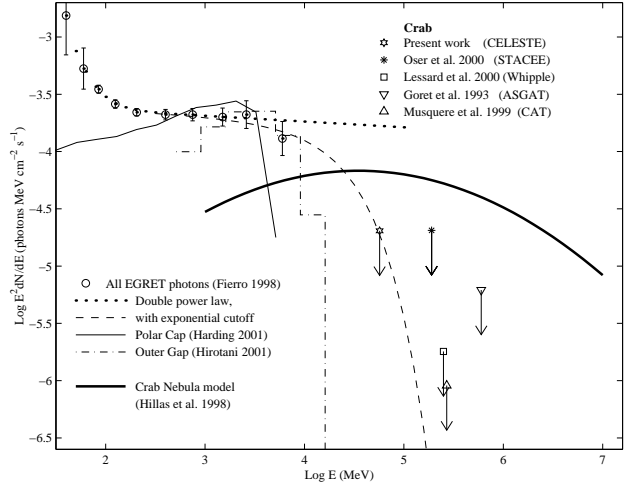}
\caption{Pulsed photon spectrum of the Crab pulsar (after 
\citet{whipplepulsed}). 
The EGRET data points are for all phase values, with
the double power-law fit from \citet{fierro}.
The fit is attenuated by
an exponential with cutoff energy $E_{0}=26\U{GeV}$ derived from the 
CELESTE upper limit (dashed curve).  
The CELESTE mean energy is shown by the open star on the curve.
The thin solid line is the polar cap model applied to the Crab (A.K.~Harding 
2001, private communication). The dotted line is the outer gap model 
corresponding to the dashed line in Figure 4 of \citet{hirotani01} (for the 
case $\mathrm{j}_{gap}=0.01,~\mathrm{j}_1=0.25,~\mathrm{j}_2=0$: see paper for 
details).  The thick solid line shows the model of unpulsed GeV-TeV emission 
from the Crab Nebula \citep{whipspec}\label{pulsedspec}}
\end{figure}

\citet{hirotani01} treat the electrodynamics of the outer gap from
first principles. The free parameter in their model is the current density 
at the gap boundaries, which in turn depends on the distance of the gap
from the light cylinder. Our upper limit excludes the hypothesis that the 
current density vanishes at the gap surface, since the model predicts
a gamma ray flux extending to 60 GeV in that case. Fig.~\ref{pulsedspec}
includes the prediction of their model for the case of a small current
density at the inner boundary, and a null current at the outer boundary.
Fig.~\ref{pulsedspec} also shows the predictions of a polar cap model, along 
with the EGRET measurements and higher energy upper limits. The CELESTE upper 
limit constrains the high energy emission more strongly than the previous 
Whipple measurement, but increased sensitivity at lower energy is still
needed to favor a particular model for the emission processes.

\section{Conclusions}

We have presented the first detection by the atmospheric \v{C}erenkov 
technique of a gamma ray source, the Crab nebula, at energies below 100\U{GeV} 
using the CELESTE experiment. The measured flux is compatible 
with most emission models. No periodic signal has been detected but our upper 
limit allows us to constrain further the cutoff point for emission from the 
pulsar. As our uncertainties decrease we will be able to determine the energy
range in which the nebula and pulsar contributions are comparable.

The data reported on in this paper were collected during the first observation 
season with a fully operational 40 heliostat array. 
In single pointing mode we now have a sensitivity to the Crab of 
$2.1\sigma/\sqrt{hour}$. A smaller dataset obtained with heliostat
double pointing appears to confirm Monte Carlo predictions of improved
sensitivity, yielding $3.4\sigma/\sqrt{hour}$, although more data is
required for confirmation.
It seems likely that \textit{a posteriori} optimization 
of our hadron rejection cuts, along with the development of new analysis 
techniques, will enable us to improve our sensitivity in the future.
CELESTE is currently being upgraded by the addition of another 
13 heliostats, bringing the total to 53, allowing greater flexibility
in pointing strategies.


\acknowledgments

Funding was provided by the IN2P3 of the French CNRS and by the Grant Agency of the Czech
Republic. We gratefully acknowledge the support of the Regional Council of
Languedoc-Roussillon and of Electricit\'e de France. We thank Dr. B. Lott, Dr. R. Lessard,
Dr. F. Aharonian, Dr A. Harding and Dr K. Hirotani for useful discussions.
Dr. F. Piron provided the CAT data used herein.

\clearpage



\clearpage

\begin{deluxetable}{ccccc}
\tabletypesize{\scriptsize}
\tablecaption{Crab observations for the 1999/2000 observing 
season.\label{tab:obslog}}
\tablewidth{0pt}
\tablehead{
\colhead{Pointing Altitude} & \colhead{Number} & \colhead{Number}   &
\colhead{ON Source Duration} & \colhead{Dates} \\
\colhead{(km)} & \colhead{of Pairs} & \colhead{Used} & \colhead{(hours)}}
\startdata
11      & 75  & 41  & 12.1 & 11/99 - 03/00 \\
11-25   & 12  & 9   & 2.2  & 01/00 - 02/00 \\
\enddata
\end{deluxetable}


\begin{deluxetable}{ccccccc}
\tabletypesize{\scriptsize}
\tablecaption{The number of events remaining at each stage of the analysis for 
the 11km single pointing Crab data set.\label{tab:crablow}}
\tablewidth{0pt}
\tablehead{
\colhead{Cut} & \colhead{Number} & \colhead{Number}   &
\colhead{Difference} & \colhead{Significance}  & \colhead{Signal/} &
\colhead{$\gamma$ rate} \\
\colhead{} & \colhead{ON} & \colhead{OFF} & \colhead{} & \colhead{($\sigma$)} &
\colhead{Background} & \colhead{(\UU{min}{-1})}} 
\startdata
Raw Trig.                    & $894\,494$ & $888\,725$ & $5\,769$ & $4.3$ & 
$0.6\%$ & \nodata \\
Software Trig.               & $474\,823$ & $469\,312$ & $5\,511$ & $5.7$ & 
$1.2\%$ & $7.6$ \\
$N_{\mathrm{peaks}}\geq 10$  & $434\,368$ & $429\,242$ & $5\,126$ & $5.5$ & 
$1.2\%$ & $7.1$ \\
$\sigma_{\mathrm{grp}}<0.25$ & $67\,022$  & $64\,295$  & $2\,727$ & $7.5$ & 
$4.2\%$ & $3.8$ \\
$\theta\leq 7\U{mrad}$       & $41\,442$  & $39\,481$  & $1\,961$ & $6.9$ & 
$5.0\%$ & $2.7$ \\
\enddata
\end{deluxetable}

\begin{deluxetable}{lrrrr} 
\tabletypesize{\scriptsize}
\tablecolumns{5} 
\tablewidth{0pt} 
\tablecaption{Cut efficiencies calculated from the real Crab 
data and from simulated gamma rays (statistical errors 
only). Shown are the incremental effects of each successive 
cut (top section), as well as the cumulative efficiencies
(bottom section). 
The last line shows the cuts used in the flux determination.\label{tab:EffAll}} 
\tablehead{ 
\colhead{}    & \colhead{}    & \multicolumn{2}{c}{Real Data} & 
\colhead{Simulation} \\ 
\cline{3-5} \\ 
\colhead{Cut} & \colhead{}   & \colhead{OFF}   & \colhead{ON-OFF} & \colhead{$\gamma$}} 
\startdata 
Software Trigger (S.T.)       &  &  $52.8 \pm 0.1\%$ &    \nodata     &  $59.4 \pm 0.3\%$ \\
$N_{\mathrm{peaks}}\geq 10$   &  &  $91.5 \pm 0.2\%$ &  $93 \pm 24\%$ &  $90.2 \pm 0.6\%$ \\
$\sigma_{\mathrm{grp}}<0.25$  &  &  $15.0 \pm 0.1\%$ &  $53 \pm 12\%$ &  $61.2 \pm 0.5\%$ \\
$\theta\leq 7\U{mrad}$        &  &  $61.4 \pm 0.4\%$ &  $72 \pm 14\%$ &  $85.2 \pm 0.7\%$ \\
\tableline
All cuts	              &  &  $4.44\pm 0.02\%$ &    \nodata     &  $28.0  \pm 0.2\%$ \\
All cuts, after S.T.          &  &  $8.41\pm 0.04\%$ &  $36 \pm 8\%$  &  $47.1  \pm 0.4\%$ \\
$N_{\mathrm{peaks}}\geq 10$, $\sigma_{\mathrm{grp}}<0.25$, after S.T.
                              &  &  $13.7\pm 0.1\%$ &  $49 \pm 17\%$  &  $55.2  \pm 0.6\%$ \\
\enddata 
\end{deluxetable} 

\begin{deluxetable}{ccccccc}
\tabletypesize{\scriptsize}
\tablecaption{The number of events remaining at each stage of the analysis for 
the $11\U{km}$ and $25\U{km}$ double pointing Crab data 
set.\label{tab:crabdouble}}
\tablewidth{0pt}
\tablehead{
\colhead{Cut} & \colhead{Number} & \colhead{Number}   &
\colhead{Difference} & \colhead{Significance}  & \colhead{Signal/} &
\colhead{$\gamma$ rate} \\
\colhead{} & \colhead{ON} & \colhead{OFF} & \colhead{} & \colhead{($\sigma$)} &
\colhead{Background} & \colhead{(\UU{min}{-1})}} 
\startdata
Raw Trig.                    & $157\,129$ & $155\,365$ & $1\,764$ & $3.2$ & 
$1.1\%$  & \nodata \\
Software Trig.               & $79\,685$  & $78\,381$  & $1\,304$ & $3.3$ & 
$1.7\%$  & $10.0$\\
$N_{\mathrm{peaks}}\geq 10$  & $75\,193$  & $73\,900$  & $1\,293$ & $3.3$ & 
$1.7\%$  & $9.9$ \\
$\sigma_{\mathrm{grp}}<0.25$ & $9\,174$   & $8\,523$   & $651$    & $4.9$ & 
$7.6\%$  & $5.0$ \\
$\theta\leq 7\U{mrad}$       & $5\,733$   & $5\,209$   & $524$    & $5.0$ & 
$10.1\%$ & $4.0$ \\
\enddata
\end{deluxetable}

\begin{deluxetable}{cc}
\tabletypesize{\scriptsize}
\tablecaption{.\label{tab:pulsed}}
\tablewidth{0pt}
\tablehead{}
\startdata
Total number of ON events  &  67022 \\
Total number of OFF events &  64295 \\
\\
Pulsed phase fraction & 0.21 \\
Number of ON events in expected phase windows	& 14062 \\
Number of ON events outside expected phase windows	& 52960 \\
Significance for the pulsed phase domain &  $-0.1\sigma$ \\
\\
Value of the H-test for ON source events &  2.60\\
Value of the H-test for OFF source events & 1.17 \\
Upper limit at the 99\% confidence level for H-test & $<31\%$ \\
\\
Upper limit at the 99\% confidence level using Helene method & $<12\%$ \\
\enddata
\end{deluxetable}




\end{document}